\documentclass[12pt,aps,nofootinbib]{revtex4}

\usepackage{graphicx}
\usepackage{amsmath,bbm}
\usepackage{amssymb,bm}
\usepackage{yfonts}

\newcommand\tr{\mathop{\mathrm{tr}}}
\newcommand\Tr{\mathop{\mathrm{Tr}}}
\newcommand\Nf{N_\mathrm{f}}

\newcommand\ai{x}
\newcommand\aii{y}
\newcommand\bi{u}
\newcommand\bii{v}
\newcommand\unitmatrix{\mathbbm{1}}
\newcommand{\pos}[1]{#1}
\newcommand\Dirac{\mathbb{D}}
\newcommand\Pmatrix{\mathbbm{P}}

\begin{document}

\author{J. Han and M.~A.~Stephanov}
\affiliation{Department of Physics, University of Illinois, Chicago, 
Illinois 60607, USA}

\title{A Random Matrix Study of the QCD Sign Problem}


\begin{abstract}

We investigate the severity of the sign problem in a random matrix
model for QCD at finite temperature $T$ and baryon chemical potential
$\mu$.  We obtain analytic expression for the average phase factor --
the measure of the severity of the sign problem at arbitrary $T$ and
$\mu$.  We observe that the sign problem becomes less severe as the
temperature is increased.  We also find the domain where the sign
problem is maximal -- the average phase factor is zero, which is
related to the pion condensation phase in the QCD with finite isospin
chemical potential. We find that, in the matrix model we studied, the
critical point is located inside the domain of the maximal sign
problem, making the point inaccessible to conventional reweighting
techniques. We observe and describe the scaling behavior of the size
and shape of the pion condensation near the chiral limit.

\end{abstract}

\maketitle 

\section{Introduction}

The phase diagram of QCD at finite temperature and baryon density has
been a subject of intense interest during recent  years (see, e.g., 
\cite{Rajagopal:2000wf}
for review).  On the one hand, the experimental heavy-ion collision
programs, such as those at CERN SPS and RHIC, as well as planned FAIR
facility, demand reliable quantitative
understanding of the phase diagram of strongly interacting matter
created in those collisions. On the other hand, deriving the relevant
predictions from the first principles of QCD is a
formidable challenge, because the phenomena associated with phase
transitions occur in the domain where the QCD coupling is not small.

One of the features of the QCD phase diagram of particular interest to
heavy-ion collision experiments is the QCD critical point.  The
existence of such a point -- an ending point of the first order chiral
transition in QCD -- was suggested a long time ago
\cite{Asakawa,Barducci}, and the properties were studied using
universality arguments and model calculations more recently
\cite{Berges,Halasz-pdqcd} (see Ref.~\cite{cp-review} for review).
The experimental search for the critical point using heavy ion
collisions has been proposed in \cite{signatures}. It is apparent that
theoretical knowledge of the location of the critical point on the
phase diagram is important for the success of the experimental search.

The time-tested approach to non-perturbative problems in
QCD is the numerical lattice Monte-Carlo calculations. This approach, very
powerful at studying QCD thermodynamics at zero baryon density, runs
into the sign-problem at nonzero baryon density. 
The lattice calculations are based on reinterpreting the QCD partition
function as a partition function of a classical statistical system,
with energy given by the Euclidean action of QCD. This action involves
the logarithm of the fermion determinant, which is complex for any
nonzero value of the baryon chemical potential $\mu_B$. The Monte Carlo
importance sampling technique uses the exponent of the action as the
measure of importance and fails because the action is complex.

Several approaches to QCD at finite baryon density are being
developed, using various techniques to circumvent or tame the
effect of the complexity of the fermion determinant and locate the
QCD critical
point~\cite{Fodor:2001au,Philipsen,Allton:2005gk,Ejiri,Gavai}
(see also Refs.~\cite{Laermann:2003cv,Schmidt:2006us,Philipsen-review,lat2006} for reviews).
As the sign problem worsens with increasing $\mu_B$, the hope is
that the QCD critical point is located at sufficiently small $\mu_B$,
where the sign problem can still be controlled.  In many cases it is
difficult to judge reliably, either {\it a priori} or {\it a posteriori}, 
what the
range of validity of the results are, in terms of how large $\mu_B$
can be before the sign problem is out of control. It is therefore 
necessary to understand better the severity of the sign problem, and
its dependence on the variables such as temperature $T$, baryon chemical
potential $\mu_B$ and quark mass $m$.

In this paper we use a random matrix model of QCD to assess the
severity of the sign problem as a function of $T$, $\mu_B$ and
$m$. A similar study at $T=0$ has been reported in
Ref.~\cite{Splittorff:2006fu,Splittorff:2007ck}. 
Here we shall present
analytical%
\footnote{A numerical study at nonzero $T$ has been reported in Ref.~\cite{Ravagli:2007rw}.}
 results for the random matrix model at {\em nonzero}
temperature -- the regime most relevant for the heavy-ion collision experiments and the lattice
studies aimed at discovering the QCD critical point.

As a quantitative measure of the severity of the
sign problem we consider the complex phase $e^{i\theta}$ 
of the fermion determinant $\det\Dirac$,
averaged over gauge field configurations of the phase-quenched theory:
\begin{equation}
  \label{eq:average-phase}
  R\equiv \langle e^{2i\theta} \rangle_{1+1^*} \equiv \left< \dfrac{\det  \Dirac^{\phantom{*}}}{\det  \Dirac^{*}}  \right>_{1+1^*} \, ,
\end{equation}
where $\Dirac$ is the  Euclidean space 
Dirac operator in a given gauge configuration: 
\begin{equation}
  \label{eq:Dirac-operator-def}
  \Dirac=\gamma^\mu(\partial_\mu - A_\mu) + m  + \mu \,\gamma_0 \, ,
\end{equation}
and $\mu$ is the quark chemical potential: $\mu=\mu_B/3$.

The average in Eq.~(\ref{eq:average-phase}), denoted by
$\langle\ldots\rangle_{1+1^*}$, is taken over the gauge field configuration
ensemble with the phase of the determinant removed (quenched), making
the measure of path integration manifestly positive: $e^{-S_{\rm
    YM}}|\det\Dirac|^2$.  This phase-quenched theory can be viewed as a
theory with 1 quark and 1 conjugate quark or, due to 
$(\det\Dirac(\mu))^*=\det\Dirac(-\mu)$, two quarks with opposite 
chemical potentials, i.e., QCD at finite isospin chemical potential
$\mu_I=2\mu$ (see, e.g,~\cite{Alford:1998sd,Son:2000xc}).
 
The average phase factor $R$ can be recast as the ratio of
two partition functions%
~\cite{Splittorff:2007ck,Ravagli:2007rw}:
\begin{equation}
  \label{eq:average-phase-ratio}
  R
\equiv \langle e^{2i\theta} \rangle_{1+1^*}
=  \dfrac{\langle \, {(\det  \Dirac)^{2} } \, \rangle_{0}}{\langle \, {\vert{\det  \Dirac }\vert^{2}} \, \rangle_{0}}=\dfrac{Z_{1+1}}{Z_{1+1^{*}}} \, ,
\end{equation}
where $\langle\ldots\rangle_{0}$ denotes average over gauge configurations in
a theory without quarks (with pure Yang-Mills measure $e^{-S_{\rm YM}}$).

\section{Random Matrix Theory (RMT) and QCD partition function}

\subsection{The random matrix model}

The Chiral Random Matrix Theory \cite{Shuryak:1992pi} approximates the
QCD partition function by an integral over random
matrix ensemble. We introduce temperature as first done in 
Ref.~\cite{Jackson:1995nf} and chemical potential as in
Ref.~\cite{Stephanov:1996ki}. The resulting random matrix model has
been used in Ref.~\cite{Halasz-pdqcd} to study the QCD phase diagram.
The partition function in the model is given by:
\begin{equation}
\label{eq:QCD-partition-function}
Z_{\Nf}=\int{\mathcal{D}X \, e^{-N\tr XX^{\dagger}}{\det}^{\Nf}\Dirac }=\langle{{\det}^{\Nf}\Dirac }\rangle_{X} \, ,
\end{equation} 
where $\Dirac$ is the $2N \times 2N$ matrix approximating the Dirac operator:
\begin{equation}
\label{eq:Dirac-operator}
\Dirac=\left( \begin{array}{ccc}
m & iX+C \\
\noalign{\medskip}
iX^{\dagger}+C & m
\end{array} \right) \, ,
\end{equation}
with
\begin{equation}
\label{eq:C-operator}
C=\mu \unitmatrix_N  + i T \left( \begin{array}{cc}
\unitmatrix_{N/2} & 0 \\
\noalign{\medskip}
0 & - \unitmatrix_{N/2}
\end{array} \right) \, .
\end{equation}  
 $X$ is an $ N \times N $ complex random matrix; $\unitmatrix_{N}$ is the $N \times N$ identity matrix.
The deterministic matrix $C$ defined by Eq.~(\ref{eq:C-operator})
accounts for the effect of the chemical potential $\mu$ and of the 2 smallest Matsubara frequencies: $+\pi T$ and $-\pi T$. For simplicity, we absorb the coefficient $\pi$ into $T$. The integration in Eq.~(\ref{eq:QCD-partition-function})
is over the real and imaginary components of the matrix~$X$:
$
\mathcal{D}X = \prod_{i,j =1}^{N} \, dX_{ij} \, dX_{ij}^{*}
$ .

All quantities appearing in the random matrix model are dimensionless, which is achieved by using the appropriate units as discussed in Ref.~\cite{Halasz-pdqcd}. The choice of the dimensionful units will not be consequential for our study.

The Dirac  determinant can be written as a Grassmann integral:
\begin{equation}
{\det}^{\Nf}\Dirac =\int{\prod_{f}^{\Nf} \mathcal{D}\psi_{R}^{f}\,\mathcal{D}\psi_{L}^{f} \, \exp\left[ \sum_{f}^{\Nf}    \left( \begin{array}{c} \psi_R^{f*}\\ \noalign{\medskip} \psi_L^{f*} \end{array} \right)^{\mathrm{T}}   \left( \begin{array}{cc} m & iX+C \\ \noalign{\medskip} iX^{\dagger}+C & m \end{array} \right)    \left( \begin{array}{c} \psi_R^{f}\\ \noalign{\medskip} \psi_L^{f} \end{array} \right)  \right] } \, ,
\end{equation}  
where the Grassmann integration is over the spinors: 
$
\mathcal{D}\psi^{f} = \prod_{i=1}^{N} \, d\psi_{i}^{f} \, d\psi_{i}^{f*}
$ .

The integration over random matrix $ X $ is Gaussian and leads to a 4-fermion interaction: $\left( \psi_{L_{i}}^{f*} \, \psi_{L_{i}}^{g} \, \psi_{R_{j}}^{g*} \, \psi_{R_{j}}^{f} \right)$ here $ i $ and $ j $ indicate dimension of random matrix $X$ while $f $ and  $g$ indicate the flavors.

Following the logic of the Hubbard-Stratonovich transformation, 4-fermion interaction can be rewritten using fermion bilinears with the help of a new auxiliary $\Nf \times \Nf $ complex matrix $A$ (flavor matrix). Performing the Grassmann integration one then obtains \cite{Halasz-pdqcd}: 
\begin{equation}
\label{eq:ZNf}
Z_{\Nf}=\int{\mathcal{D}A\, e^{(-N \tr AA^{\dagger})} \, {\det}^{\frac{N}{2}} \left( \begin{array}{cc} A+m & \mu + i T \\ \noalign{\medskip} \mu + i T & A^{\dagger}+m \end{array} \right)\,{\det}^{\frac{N}{2}} \left[ \: T \rightarrow -T \: \right] } \, .
\end{equation}
We shall specialize to $\Nf=2$ quark flavors.
The integral in Eq.~(\ref{eq:ZNf}) is performed over $2 \times
\Nf \times \Nf = 8$ variables which are real and imaginary parts of the
elements of the complex $\Nf \times \Nf = 2 \times 2$ flavor matrix $A$.
We shall define  potential $\Omega_{1+1}(A)$ as
\begin{equation}
  \label{eq:omega-1plus1}
  Z_{1+1} \equiv \int{\mathcal{D}A\, e^{-N \Omega_{1+1} (A)}},
\end{equation}
i.e.,
\begin{equation}
\label{eq:Omega1plus1}
\Omega_{1+1}(A) = \Tr \, \left[AA^{\dagger}-\frac{1}{2} \ln \left\{[(A+m)(A^{\dagger}+m)-(\mu+i  T)^{2}]\times  \left[ \: T \rightarrow -T \: \right]  \right\}\right] \, . 
\end{equation}

\subsection{Phase quenched partition function} 

The phase-quenched partition function is given by
\begin{equation}
\label{eq:Z1plus1star}
Z_{1+1^{*}}=\int{\mathcal{D}X \, e^{-N\,\tr XX^{\dagger}}{\det}\Dirac  \,\, {\det}\Dirac ^{*}} = \langle \, \vert {\det}\Dirac  \vert^{2} \, \rangle_{X} \, .
\end{equation}

Following the steps outlined above, the phase quenched partition function
$ Z_{1+1^{*}} $ can be written as
\begin{align}
\label{eq:Z1plus1star-flavor}
Z_{1+1^{*}} &= \int{\mathcal{D}A\, e^{(-N \tr AA^{\dagger})} \, {\det}^{\frac{N}{2}} \left( \begin{array}{cc} A+m & \mu\tau_3 + i T \\ \noalign{\medskip} \mu\tau_3 + i T & A^{\dagger}+m \end{array} \right)\,{\det}^{\frac{N}{2}} \left[ \: T \rightarrow -T \: \right] }\\
&\equiv \int{\mathcal{D}A\, e^{-N \Omega_{1+1^{*}}(A)}}.
\end{align}
Here $\tau_{3}$ 
is the Pauli matrix, and we defined another
potential  $\Omega_{1+1^{*}}(A)$ --- a function of a $2 \times 2$ complex flavor matrix $A$.
For a generic matrix $A$, $\Omega_{1+1^{*}}(A)\neq\Omega_{1+1}(A)$,
due to the presence of the Pauli matrix in Eq.~(\ref{eq:Z1plus1star-flavor}).

\subsection{Thermodynamic limit $N \rightarrow \infty $ and the
  solution of the model}

In the thermodynamic limit $ N \rightarrow \infty $, the random matrix partition
function $Z_{1+1}$ can be calculated analytically using the saddle-point
approximation by minimizing $\Omega_{1+1}(A)$ with respect to $A$. 
The minimum is given by a multiple of the unit matrix:
$A=a\unitmatrix$ with $a$ -- real.

For the $1+1^*$ theory the minimum of $\Omega_{1+1^*}(A)$ is also
given by a multiple of the unit matrix, except for a region on the
phase diagram where another, deeper minimum is given by a non-diagonal
matrix $A$~\cite{Stephanov:1996ki}. This breaks the U(1) ($\tau_3$ 
isospin)
symmetry of the theory in this region and is associated with the pion
condensation in QCD.

Outside of the pion condensation region, i.e., when the minimum of
$\Omega_{1+1^*}$ is given by a multiple of the unit matrix, the
minimum values of the two potentials coincide. Indeed, for any real~$a$,
$\Omega_{1+1}(a\unitmatrix) = \Omega_{1+1^*}(a\unitmatrix) \equiv
\Omega(a) $, where we defined
\begin{equation}
\label{eq:common-potential}
\Omega(a) = 2 a^2 - \ln \{ [(a+m)^2 - (i T+\mu)^2] [(a+m)^2 - (i T-\mu)^2] \} \, . 
\end{equation}
Then the saddle-point value of $A=a\unitmatrix$ for both $Z_{1+1} $ and $Z_{1+1^*} $ is determined by minimizing the potential $\Omega(a)$ with respect to $a$: 
\begin{equation}
\label{eq:saddle-point-eqn}
a - \dfrac{(a+m)[(a+m)^{2}-\mu^{2}+T^{2}]}{[(a+m)^{2}-\mu^{2}+T^{2}]^{2}+4\mu^{2}T^{2}} = 0 \, .
\end{equation}


\section{Average Phase Factor in Phase Quenched Theory}

\newcommand\Asp{A_{\rm sp}}

\subsection{ General result: arbitrary $ m, \, \mu $  and  $T $ }

Since the leading exponential behavior of the partition functions
$Z_{1+1}$ and $Z_{1+1^*}$ is the same (see previous section), it will
cancel in the ratio $R$. Therefore, we have to take into account the
preexponential factors, which are determined by the second order derivatives of the potential function $\Omega_{1+1}(A)$ and $\Omega_{1+1^{*}}(A)$ with respect to all elements of flavor matrix~$A$: 
\begin{equation}\label{eq:Z-saddle-point}
Z_{Q} \stackrel{N\to\infty}{\to} 
\left( \frac{2 \pi}{N}\right)^4 
\left(  \det \Omega_{Q}^{''} \right)^{-\frac{1}{2}} 
e^{-N \Omega_{Q}({A})} \Bigg|_{A=\Asp} \, ,
\end{equation}
where $Q$ indicates the quark content of the theory, $1+1$ or $1+1^*$, and
\begin{equation}
  \label{eq:det-2prime}
  \det \Omega_{Q}^{''}\equiv 
\det\left(\frac{\partial^{2}\Omega_{Q}}{\partial A_{\alpha} \partial A_{\beta}}\right) \, .
\end{equation}
The indices $\alpha$ and $\beta$ run through eight values labeling
eight independent components of the complex $2\times2$ matrix $A$:
$A_{\alpha} $ and $A_{\beta}$ = $(A_{11}, A_{11}^{*},A_{12},
A_{12}^{*},A_{21}, A_{21}^{*},A_{22}, A_{22}^{*} )$.
Evaluating determinants in Eq.~(\ref{eq:det-2prime}) at the saddle point
$A=\Asp$ we find, using notations given below in Eq.~(\ref{eq:factors}):
\begin{equation}
\label{eq:determinants}
\det \Omega_{1+1}^{''} = [\ai^2-\aii^2]^4  
\quad \mbox{and} \quad 
\det \Omega_{1+1^*}^{''} = [\ai^2-\aii^2]^2 [\bi^2-\bii^2]^2 \, .
\end{equation}   

Since $\Omega_{1+1}(\Asp)=\Omega_{1+1^*}(\Asp)$ outside of the region
of pion condensation in $1+1^*$ theory, the exponential factors in
Eq.~(\ref{eq:Z-saddle-point}) cancel in the ratio $R$ and the
average phase factor is given by
\begin{equation}
\label{eq:pre-exponential-factors}
R \equiv \langle{e^{2i\theta}}\rangle_{1+1^{*}} =\dfrac{Z_{1+1}}{Z_{1+1^{*}}} =\left[ \dfrac{\det \Omega_{1+1}^{''}}{\det \Omega_{1+1^*}^{''}}\right]^{- \frac{1}{2}}  =  \dfrac{\bi^{2}-\bii^{2}}{\ai^{2}-\aii^{2}} \Bigg|_{A=\Asp} \, ,
\end{equation}
where we define $ \ai, \aii, \bi, \bii $ as follows:

\begin{align}
\label{eq:factors}
\ai  &= 1 - \dfrac{T^{2}-\mu^{2}}{W} - \dfrac{8T^{2}\mu^{2}(a+m)^{2}}{W^{2}} \, ; \nonumber \\
\aii  &= \dfrac{(a+m)^{2}}{W} \left( 1- \dfrac{8T^{2}\mu^{2}}{W} \right) \, ; \nonumber  \\
\bi  &= 1 - \dfrac{T^{2}+\mu^{2}}{W} \, ; \\
\bii  &=  \dfrac{(a+m)^{2}}{W} \, ; \nonumber \\
W  &= [(a+m)^{2} + T^{2} - \mu^{2}]^{2}+4\mu^{2}T^{2} \, . 
\nonumber
\end{align}
where $a$ is a solution of the saddle-point equation~(\ref{eq:saddle-point-eqn}), and the global minimum of $\Omega(a)$ in Eq.~(\ref{eq:common-potential}).

\begin{figure}
\includegraphics[width=0.8\columnwidth]{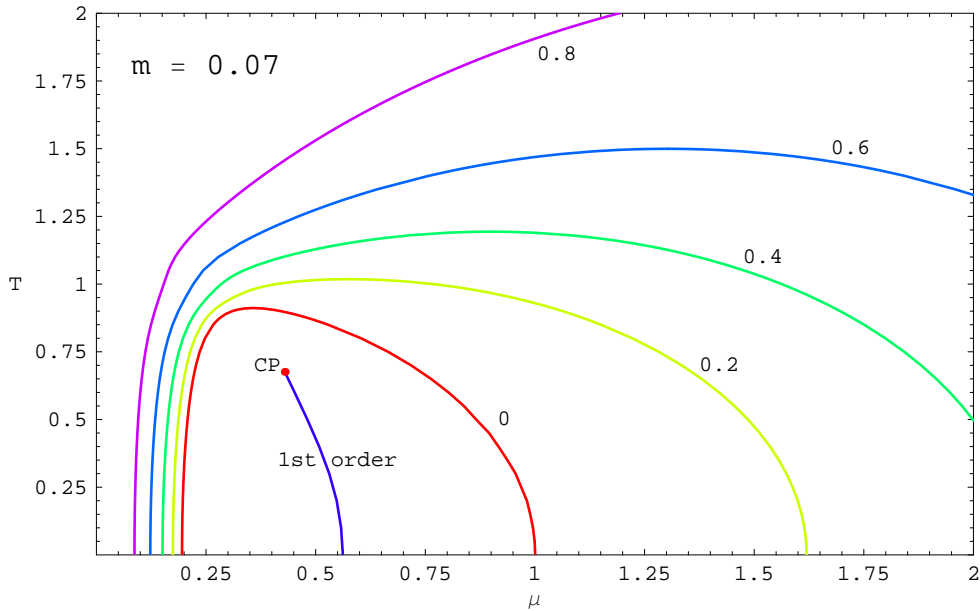} 
\caption{The contours of the average phase factor $R$ for
  $m=0.07$. The first order phase transition line in $1+1$ theory and
  the critical point are also shown (see discussion in Section~\ref{sec:qcd-sign-problem}).}
\label{fig:contours_m007}
\end{figure}

Using Eqs.~(\ref{eq:pre-exponential-factors}) and~(\ref{eq:factors})
we can now obtain the average phase factor $R$ for any values of $T$, $\mu$ and $m$. In general, this has to be done numerically, but in certain limiting cases, discussed below, explicit analytical results can be derived as well.
As an illustration, the average phase factor contours on the $T\mu$ plane for $m=0.07$ are plotted in Figure \ref{fig:contours_m007}. 

\subsection{Arbitrary $ T $ and $ \mu $  in the chiral limit $m=0$ }

In the chiral limit $m=0$ the solution of the saddle-point equation~(\ref{eq:saddle-point-eqn})
can be found explicitly. In the high-temperature phase it is simply $a=0$.
Then, from (\ref{eq:pre-exponential-factors}) and (\ref{eq:factors}) the average phase factor $R$ can be written explicitly: 
\begin{equation}
\label{eq:average-phase-zero-m}
R \equiv \langle{e^{2i\theta}}\rangle_{1+1^{*}} =\dfrac{[(T^{2}+\mu^{2})^{2}-(T^{2}+\mu^{2})]^{2}}{[(T^{2}+\mu^{2})^{2}-(T^{2}-\mu^{2})]^{2}}
\qquad (m=0)\,. 
\end{equation}
From 
(\ref{eq:average-phase-zero-m}), the contour where the average phase factor vanishes is determined by: $ T^{2}+\mu^{2} = 1$. 

\begin{figure}
\includegraphics[width=0.8\columnwidth]{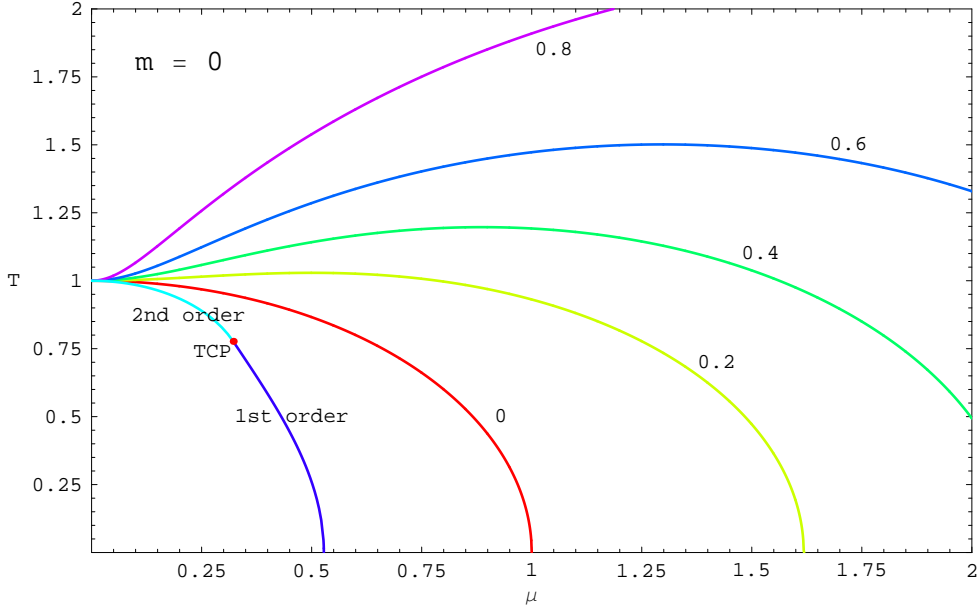} 
\caption{The contours of the average phase factor $R$ in the chiral limit
  $m=0$. The chiral symmetry transition line and tri-critical point (TCP) are also shown.} \label{fig:contours_zero_mass}
\end{figure}

The contour plot of $R(T,\mu)$ at $m=0$ is shown in Figure~\ref{fig:contours_zero_mass}.

\subsection{Small $m$ and $\mu$}


At small $m$ and $\mu$ the solution to the fifth-order polynomial equation
for the saddle point $a$ can be found in the form of an expansion, which
at zero temperature is given by
\begin{equation}
\label{eq:A-zero-T}
a \approx 1-\dfrac{m}{2}+\dfrac{\mu^{2}}{2}
\qquad   (T=0,\quad m\sim\mu^2\ll1) \, ,
\end{equation}
and thus
\begin{equation}
\label{eq:average-phase-zero}
R \approx 1-\dfrac{2{\mu}^{2}}{{m}}
=1-\dfrac{4\mu^{2}}{m_{\pi}^{2}}
\qquad   (T=0,\quad m\sim\mu^2\ll1) \, ,
\end{equation}
where we used the fact that the $R=0$ value is achieved at the
boundary of the pion condensation phase in $1+1^*$ theory, and that,
in QCD, the pion condensation occurs at $\mu_u=-\mu_d=m_\pi/2$ to
define the pion mass: $m_\pi^2=m/2$ in the units employed in the
random matrix model. This result is in agreement with the earlier $T=0$
calculation~\cite{Splittorff:2006fu,Splittorff:2007ck}.


Now, using our general result, we can extend the result of 
Ref.~\cite{Splittorff:2006fu,Splittorff:2007ck} to nonzero temperatures.
In this case, the saddle point is given by: 
\begin{equation}
\label{eq:A-small-T}
a \approx \sqrt{1-T^{2}}-\dfrac{m}{2}
\left(\dfrac{1-2T^{2}}{1-T^{2}}\right) +\dfrac{\mu^{2}}{2}
\left(\dfrac{1-4T^{2}}{\sqrt{1-T^{2}}}\right)  
\qquad(T<1,\quad m\sim\mu^2\ll1) \, ,
\end{equation}

and 
\begin{equation}
\label{eq:average-phase-small}
R\approx 1-\sqrt{1-T^{2}}\,\left(\dfrac{2 \mu^{2}}{m}\right)=1-\sqrt{1-T^{2}}\,\left(\dfrac{4\mu^{2}}{m_{\pi}^{2}}\right)
\qquad(T<1,\quad m\sim\mu^2\ll1) \, ,
\end{equation}
which generalizes Eq.~(\ref{eq:average-phase-zero}). Equation~(\ref{eq:average-phase-small})
shows that the sign problem diminishes at higher temperatures, which
can be seen also on the contour plot in Figure~\ref{fig:contours_m007}.

\subsection{$R=0$ contour}

The contour where $R$ vanishes is of particular interest to our study --
the sign problem reaches its maximum there, i.e., the fluctuations
of the phase completely wash out the magnitude.

We can obtain an explicit equation for the $R=0$ contour 
by setting $ \bi = \bii $ in equations~(\ref{eq:pre-exponential-factors}), (\ref{eq:factors}):
\begin{equation}
\label{eq:A-eqn-zero-contour}
\bi = \bii \quad \Longrightarrow \quad [(a+m)^2 + T^2-\mu^2] m =2 \mu^2 a \, .
\end{equation}
Solving the quadratic equation (\ref{eq:A-eqn-zero-contour}) for $a$ and
substituting the solution into the saddle-point equation
(\ref{eq:saddle-point-eqn}) one finds for the $R=0$ contour: 
\begin{equation}
\label{eq:zer-contour-eqn}
T^2 = 1 -\mu^2 +\frac{m^2}{\mu^2-m^2}-\frac{m^2}{4(\mu^2-m^2)^2} \, ,
\end{equation}
in agreement with an earlier result \cite{Vanderheyden:2001gx}.

It is interesting to consider the limiting behavior (shape) of this
contour as $m\to0$. As can be seen from Figures \ref{fig:contours_m007} and \ref{fig:contours_zero_mass}
the contour develops a singularity (a kink) at $\mu=0$, $T=1$ in this limit.
Near this kink the shape and location of the contour scale with $m$,
i.e., the contour at different values of $m$ can be obtained by
rescaling 
\begin{equation}
  \label{eq:scaling}
  t\to\lambda^{2/3}t \quad\mbox{and}\quad 
  \mu^2\to\lambda^{2/3}\mu^2 \quad\mbox{as}\quad 
  m\to \lambda m  \, ,
\end{equation}
where we introduced $t\equiv T^2-1$.
This can be seen upon expanding Eq.~(\ref{eq:zer-contour-eqn}) in
$\mu$, $t$ and $m$ in the regime $t:\mu^2:m^{2/3}$ fixed as $m\to0$:
\begin{equation}
  \label{eq:R0-contour-mto0}
  t = -\mu^2 -\frac{m^2}{4\mu^4} + {\cal O}(m^{4/3}) \, .
\end{equation}

For example, the point where the $R=0$ contour  reaches maximum
temperature slides to $T\to1$ and $\mu \to 0$ as
\begin{equation}
  \label{eq:max-T-point}
  T_{*}^2 = 1-\dfrac{3}{2^{4/3}} \, m^{2/3} + {\cal O}(m^{4/3})
\qquad\mbox{and}\qquad
\mu_{*}^2 = \dfrac1{2^{1/3}}\,{m^{2/3}} + {\cal O}(m^{4/3}) \, .
\end{equation}

This result may be useful for analysis of QCD simulations using the
improved reweighting~\cite{Fodor:2001au}
 or techniques which treat the sign problem
by separating the phase from the absolute value of the determinant.
The reweighting methods break down at $R=0$, and the problem is to
distinguish the signatures of the critical point, which are
similar to those of the breakdown of the
reweighting~\cite{Splittorff:2005wc,Ejiri:2005ts}. By doing
simulations at different values of $m$ and comparing to scaling
(\ref{eq:scaling}) one can determine whether the observed signatures
are those of the genuine critical point of $1+1$ theory or those of 
the breakdown of the
reweighting
method.

To be precise, in QCD, the scaling behavior of the phase transition
boundary in the phase quenched, $1+1^*$ theory (i.e., pion
condensation at finite isospin chemical potential) should be similar
to (\ref{eq:scaling}) , but with critical exponent $1/(\beta\delta)$
replacing the mean-field exponent $2/3$.  The value
$1/(\beta\delta)\approx 0.54$ is the ratio of critical scaling
dimensions of the energy-like and ordering-field-like operators in the
$O(4)$ universality class of the QCD phase transition (for two
flavors). The scaling shape of the pion condensation boundary, given
by Eq.~(\ref{eq:R0-contour-mto0}) in the random matrix model, in QCD
will also be correspondingly different.

\subsection{$R=0$ domain and the sign problem}

\begin{figure}
  \centering
\includegraphics[height=0.4\textheight]{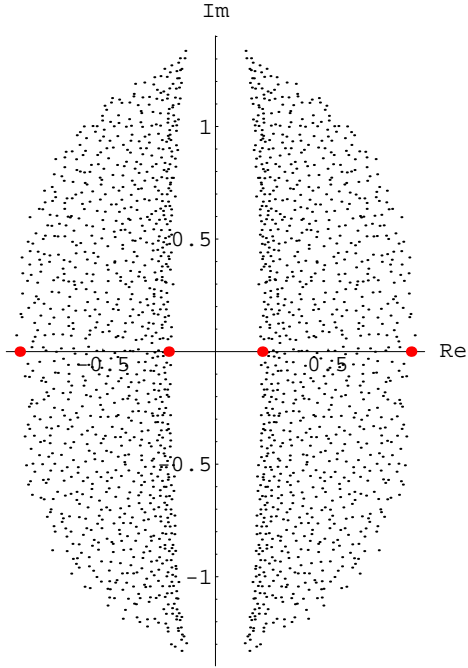}
\caption[]{ The distribution of zeros of $\det\Dirac$ 
  in the complex $\mu$ plane for a random
  matrix model with finite $N=1000$ at $T=0.5$ and $m=0.07$.  The
  filled circles indicate solutions of the $R=0$ contour
  equation~(\ref{eq:zer-contour-eqn}), bounding the region of the
  maximal sign problem.}
\label{fig:zeros}
\end{figure}

One can understand the underlying reason that the sign problem becomes severe in
the $R=0$ region by looking at the distribution of the zeros of the
$\det\Dirac$ as a function of $\mu$. These zeros can be also viewed as
eigenvalues of the random matrix $\Pmatrix$, defined as%
\footnote{This is the analog of the propagator matrix introduced in \cite{Gibbs:1986hi}.}
\begin{equation}
  \label{eq:P-matrix}
\Dirac = (\mu\unitmatrix_{2N} - \Pmatrix)\gamma_0 \, ,
\qquad\mbox{where}\quad
\gamma_0=
\left( 
\begin{array}{cc}
0&\unitmatrix_N\\\unitmatrix_N&0\\
\end{array}
\right) \, .
\end{equation}
The locations of the zeros are random, fluctuating together with matrix
$X$ in the ensemble in Eq.~(\ref{eq:QCD-partition-function}). For
$N\to\infty$ the density of zeros develops finite region of
support with a sharp boundary, which one can see already quite clearly
for a finite, but large, matrix on Figure~\ref{fig:zeros}.%
\footnote{At $T=0$ these distributions have been studied in Ref.~\cite{Halasz:1999gc}.}

The fluctuations of the phase
of $\det\Dirac$ become large when $\mu$ enters the domain of the
zeros. More explicitly, one can write
\begin{equation}
  \label{eq:detD}
  \arg\det\Dirac = \sum_i \arg(\mu - \lambda_i) \, ,
\end{equation}
where the sum is over all eigenvalues $\lambda_i$ of the matrix
$\Pmatrix$.  When $\mu$ is away from the domain of support of the
eigenvalue density, the fluctuations in the eigenvalue positions
do not affect the phase of the determinant significantly. On the other
hand, when $\mu$ is in the domain filled with eigenvalues, there are
eigenvalues which come very close to $\mu$ (as close as $1/\sqrt
N$) and even small fluctuations in the eigenvalue positions translate
into large fluctuations of the phase, causing severe sign problem.

To confirm this picture, one can compare the solutions of
the equation~(\ref{eq:zer-contour-eqn}) with the distribution of the
eigenvalues at the same values of $T$ and $m$, 
as it is shown in Figure~\ref{fig:zeros}. The comparison clearly shows
that $R=0$ (the sign problem is most severe) for the values of $\mu$
which fall inside the domain of eigenvalues.

\subsection{QCD sign problem vs RMT}
\label{sec:qcd-sign-problem}

When drawing conclusions from our random matrix study for QCD one
should bear in mind the following. In QCD, in thermodynamic
$V\to\infty$ limit,
the partition functions $Z_{1+1}$ and $Z_{1+1^*}$ are already different
at the exponential level. Indeed, expressing the partition function
via the pressure, $Z=\exp(VP(T,\mu))$, one finds that%
\footnote{For simplicity, we denote by $V$ the 4-dimensional
  Euclidean volume
  of the finite-temperature system: $V\equiv V_{\rm space}/T$.}
\begin{equation}
  \label{eq:pressures}
  R = \frac{Z_{1+1}}{Z_{1+1^*}}=\exp[V(P_{1+1}-P_{1+1^*})] \, .
\end{equation}
As an illustration, consider sufficiently low temperatures ($T\ll m_\pi)$.
With exponential precision, the $\mu$-dependence of pressure is given by the masses of the lightest particles
with nonzero charge to which the chemical potential couples, i.e.,
$P_{1+1}(T,\mu)-P_{1+1}(T,0)\sim \mu^2e^{-m_N/(3T)}$ and 
$P_{1+1^*}(T,\mu)-P_{1+1^*}(T,0)\sim \mu^2e^{-m_\pi/(2T)}$, as long as $T\ll m_\pi$. Since
$P_{1+1}(T,0)=P_{1+1*}(T,0)$, and neglecting $e^{-m_N/(3T)}$ compared
to $e^{-m_\pi/(2T)}$ we can write (with double exponential precision):
\begin{equation}
  R(T,\mu) = \frac{Z_{1+1}(T,\mu)}{Z_{1+1^*}(T,\mu)} 
  \sim \exp[-V\mu^2e^{-m_\pi/(2T)}] \, ,
\end{equation}
which means for $T\ll m_\pi$ the phase factor $R$ is exponentially
small for large $V$.~\footnote{Interestingly, using the condition
$R<1$ one could derive the mass inequality: $m_\pi/2<m_N/3$~\cite{Cohen:2003ut}.}
 This is not surprising, since the ``warm'' gas of
pions is very much different from the gas of baryons at the same
temperature.%
\footnote{Of course, as $T\to0$, the ratio $R\to1$,
which is the reflection of the fact that for $T=0$ there is no dependence on any chemical potential (for $\mu<m_\pi/2$), 
neither baryon nor isospin.}

In contrast, in the random matrix model, as we have seen, the
exponential of the volume (i.e., the matrix size $N$) cancels in the
ratio $R$.

However, in QCD, for $T\sim T_c$ and higher, the difference between
the pressures, $P_{1+1}(T,\mu)-P_{1+1^*}(T,\mu)$, although remaining
of order $V\mu^2$, becomes smaller. Lattice studies indicate
remarkable similarity of the partition functions $Z_{1+1}$ and
$Z_{1+1^*}$, manifested, e.g., in the similar slopes $dT_c/d\mu^2$ of
the pseudo-critical lines, and it has been also argued that the
difference between the slopes is suppressed in the large $N_c$ limit
of QCD~\cite{Toublan:2005rq}. It may be also added that in QCD at
asymptotically large $T\gg T_c$ the difference vanishes due to the
asymptotic freedom -- quark flavors decouple from each other, and the
pressure is independent of the relative sign of the quark chemical
potentials.  Thus the random matrix model results might be a useful
guide to the sign problem in QCD at least in the range of temperatures
near $T_c$, which is of much experimental and theoretical interest.

We wish to stress again that for the comparison of QCD sign problem to
the RMT to be meaningful the lattice 4-volume $V$ should be {\em
  finite}. More quantitatively, the system should be within the
so-called epsilon-regime~\cite{Splittorff:2006vj,Gasser:1987ah}, 
where the RMT description of the QCD becomes exact.
In QCD, as the volume is increased, the role of the exponential factor
eventually becomes dominant and $R$ vanishes exponentially. The
crossover between the epsilon-regime and the thermodynamic limit can be
studied, e.g., along the lines of Ref.~\cite{Splittorff:2007zh}.

With the preceding discussion in mind, let us 
take the point of view~\cite{Splittorff:2005wc} that the sign problem
becomes intractable (even on a finite volume), when we enter the domain
of pion condensation in the phase-quenched theory, i.e.,
$R\equiv\langle{e^{2i\theta}}\rangle=0$ domain of the RMT. In this context, it is
interesting to see where, relative to this domain, is the critical
point of the $1+1$ theory. In the 2-flavor random matrix model we
studied, the critical point, as well as the whole first order
transition line, is always inside the $R=0$ domain, as
Figures~\ref{fig:contours_m007} and
\ref{fig:contours_zero_mass} demonstrate.

\section{Summary and conclusions}

We investigated the strength of the QCD sign problem and its
dependence on temperature $T$ and quark mass $m$ using the random
matrix model.  We observed that the sign problem diminishes at higher
temperatures, which is a welcomed property, anticipated, e.g., in
\cite{Alford:1998sd}, and relied upon in improved reweighting techniques,
e.g., Ref.\cite{Fodor:2001au}.

We also observed that the strength of the sign problem is related to
the position of the pion-condensation region in the phase-quenched
theory, equivalent to the theory where the baryon chemical potential
is replaced by the isospin chemical potential (of the same absolute
value per quark), as already discussed in
Ref.~\cite{Splittorff:2005wc}. In particular, this underscores the
importance of understanding the phase diagram of QCD at finite
isospin chemical potential~\cite{Alford:1998sd,Son:2000xc,Kogut:2004zg,Sinclair:2006zm,deForcrand:2007uz}.

We observed and generalized to QCD the scaling behavior of the shape of
the pion condensation region in the $T\mu$ plane near $T=T_c$ as
$m \to 0$, where it develops a singularity (kink). This allows us
to understand how the phase diagram of QCD with finite isospin
chemical potential evolves towards the chiral limit. One of the
practical applications of this result is to guard the reweighting
techniques against a possible breakdown of the reweighting method
by studying the dependence of the breakdown point on the quark mass.

In the random matrix model we studied, the critical point of the
$1+1$ flavor theory falls within the $R=0$ domain of the maximal
sign problem (see Figures~\ref{fig:contours_m007}
and~\ref{fig:contours_zero_mass}). This means that reweighting methods
cannot access the critical point in such a theory. However, it is
possible, that a 3-flavor theory, like QCD with a strange quark, where
the critical point lies at smaller values of $\mu$ (tunable by the
strange quark mass), that the critical point is outside the pion condensation domain. What happens in this case is an 
interesting question which we hope to address in future work.

\subsection*{Acknowledgements}

The authors thank P.~de~Forcrand, K.~Splittorff and J.~Verbaarschot for  valuable comments.
This research is supported by the DOE grant No.\ DE-FG0201ER41195.




\end{document}